\documentclass[aps,a4paper,superscriptaddress,twocolumn,showpacs,preprintnumbers,amsmath,amssymb]{revtex4}
\usepackage{graphicx}
\usepackage{bm}

\newcommand{\degree}{^\circ}

\providecommand{\ep}{$e^{\scriptscriptstyle-}\!/\!p^{\scriptscriptstyle+}$}
\newcommand{\RB}{\textbf{R}$\times$\textbf{B}{} }
\begin{document}
	
	\title{Spectroscopy with cold and ultra-cold neutrons}
	
	\author{Hartmut Abele}\email{abele@ati.ac.at}
	\author{Tobias Jenke}\email{tjenke@ati.ac.at}
	\author{Gertrud Konrad}\email{gkonrad@ati.ac.at}
	
	\affiliation{Atominstitut, Technische Universit\"at Wien, Stadionallee 2, A-1020 Wien, Austria}
	
	\date{\today}
	
\begin{abstract}
We present two new types of spectroscopy methods for cold and ultra-cold neutrons.  The first method, which uses the \RB drift effect to disperse charged particles in a uniformly curved magnetic field, allows to study neutron $\beta$-decay.  We aim for a precision on the 10$^{-4}$ level.  The second method that we refer to as gravity resonance spectroscopy (GRS) allows to test Newton's gravity law at short distances.  At the level of precision we are able to provide constraints on any possible gravity-like interaction.  In particular, limits on dark energy chameleon fields are improved by several orders of magnitude.
\end{abstract}
\maketitle
\section{Introduction}
\label{intro}
Neutrons react to all known forces and are a powerful tool for addressing fascinating questions in particle physics, nuclear physics, and astronomy.  It belongs to the opportunities that the investigation of static and decay properties of the free neutron are key issues in particle physics and astrophysics, which can be addressed complementary to the high-energy physics approach.  Precision studies of Newton's law at very small distances in turn allow to probe for extra dimensions at the $\mu$m level and can reveal the existence of new gauge bosons acting within.

Precise symmetry tests of various kinds are coming within reach with the proposed facility PERC~\cite{Dub08,Kon12}.  Projects using the PERC facility will test the Standard Model at a much higher level of sensitivity benefiting both, from the gain in statistical accuracy for individual measurements and from the redundancy of observables accessible.  Neutron decay offers a number of independent observables, considerably larger than the small number of parameters describing this decay in the Standard Model.  Examples are the electron-antineutrino correlation coefficient $a$~\cite{Str77,Byr02,Baessler2008,Kon09}, the beta asymmetry parameter $A$~\cite{Abele97,Abele02,Mund13,Plaster12,Mendenhall13}, the neutrino asymmetry parameter $B$~\cite{Kreuz05a,Schumann07} (reconstructed from proton and electron momenta), the proton asymmetry parameter $C$~\cite{Schumann08a}, the triple correlation coefficient $D$~\cite{Soldner04,Chupp12}, the Fierz interference term $b$, and various correlation coefficients involving the electron spin~\cite{Kozela09,Kozela12}.  Each coefficient in turn relates to an underlying broken symmetry.  A method of loss-free spectroscopy is presented in Ref.~\cite{Abele1993}.

In Sec.~\ref{sec-1}, we present a novel spectroscopy technique for electron and proton spectroscopy, which can be used with PERC.  In Sec.~\ref{sec-2}, we present the first precision measurements of gravitational quantum states with GRS.

\section{Neutron $\beta$-decay and \RB spectroscopy}\label{sec-1}

The facility PERC (Proton and Electron Radiation Channel)~\cite{Dub08}, for high-precision measurements of neutron $\beta$-decay, is under development~\cite{Kon12}.  The basic idea of PERC is to supply its users with an intense beam of well-defined electrons and protons (\ep) from free neutron decay.  The all-purpose \ep-beam allows to measure a variety of neutron decay observables related to physics in and beyond the Standard Model~\cite{Jac57,Her01,Sev06,Abe08,Dubbers11,Severijns11}.

\begin{figure}[htb]
	\centering
	\includegraphics[width=0.48\textwidth]{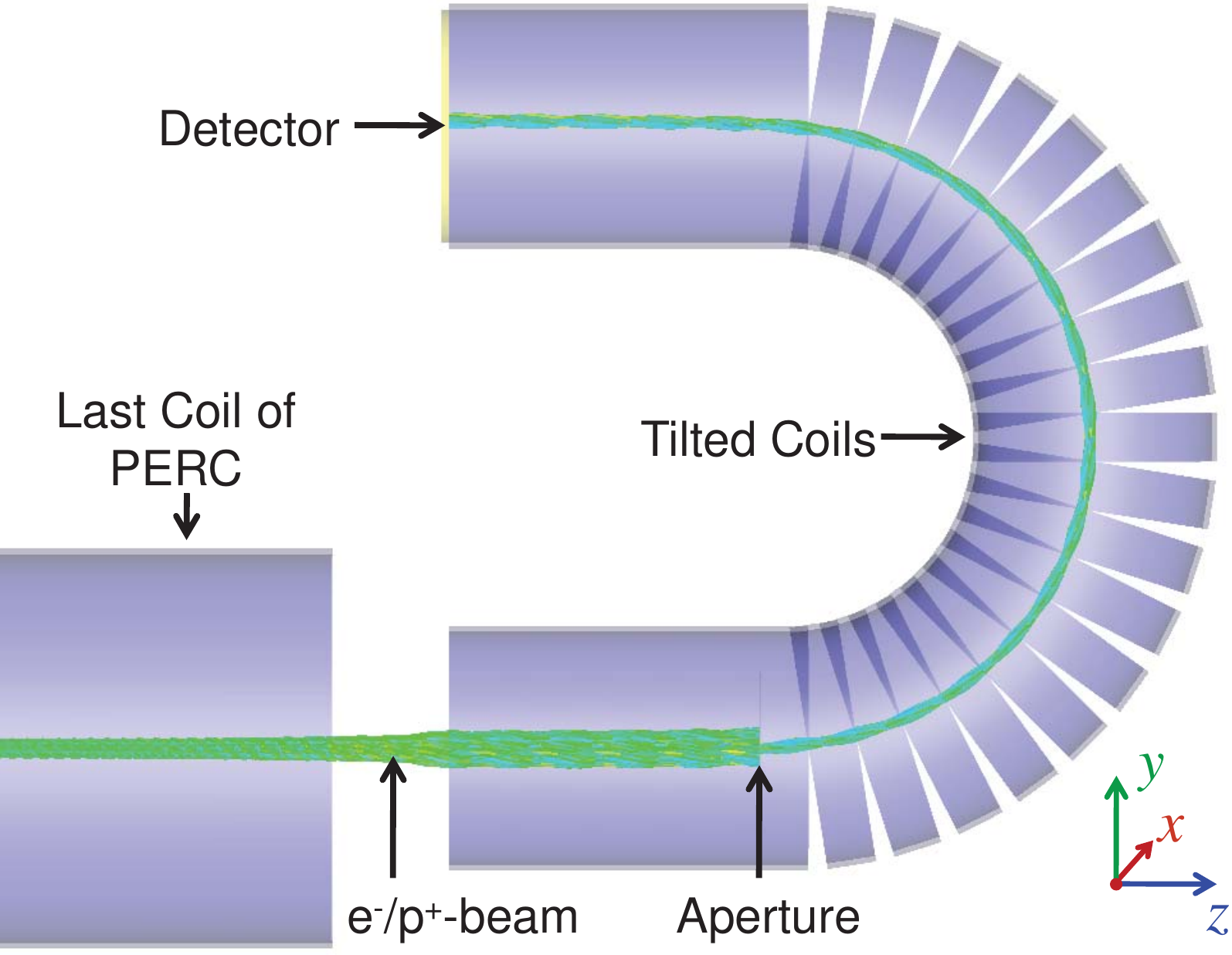}
	\caption{Scheme of the \RB drift spectrometer~\cite{Wang13} connected to the end of PERC, with simulated \ep-trajectories in green.}
	\label{figure:driftspec geometry1}
\end{figure}

Cold neutrons pass through the decay volume of PERC where only a small fraction decays into charged {\ep} and neutral electron antineutrinos.  The charged {\ep} are guided by the strong magnetic field of PERC towards a user's detection system.  Figure~\ref{figure:driftspec geometry1} shows as an example the \RB drift spectrometer connected to the end of PERC.

High momentum resolution is provided by magnetic spectrometers.  The resolution $\Delta p = eB \cdot \Delta r$ for momentum $p$ of a charge $e$ in a magnetic field $B$ depends on the resolution for the radius $r$ of the spatial resolution detector.

The \RB drift spectrometer~\cite{Wang13} provides the opportunity to measure the momentum of charged particles in the presence of a guiding field, in which case normal magnetic spectrometers cannot work well:  Instead of eliminating the guiding field, it is gradually adapted to the analysing, curved magnetic field.  In the uniformly curved magnetic field, the drifts of the particles have similar behaviours as in normal magnetic spectrometers.

The advantages of the \RB drift spectrometer over normal magnetic spectrometers are:
\begin{itemize}
	\item \emph{Adiabatic transport of particles:}
	As shown in Fig.~\ref{figure:driftspec geometry1}, charged particles are adiabatically transported from the guiding field in front of to the detector at the end of the \RB drift spectrometer.  In this way, the angular distribution of the particles can be preserved and measured.
	\item \emph{Low momentum measurements:}
	Particles with infinitely small momentum $p\rightarrow 0$ can be measured, whereas they cannot be measured with normal magnetic spectrometers if their dispersion is smaller than the aperture width.
	\item \emph{Large acceptance of incident angle:} 	
	For incident angles smaller than 10$^{\degree}$ the aberration is smaller than 10$^{-4}$.
\end{itemize}

\section{Gravity Resonance Spectrometry constrains dark matter and dark energy\label{sec-2}}

\begin{figure}[htb]
	\includegraphics[width=3cm]{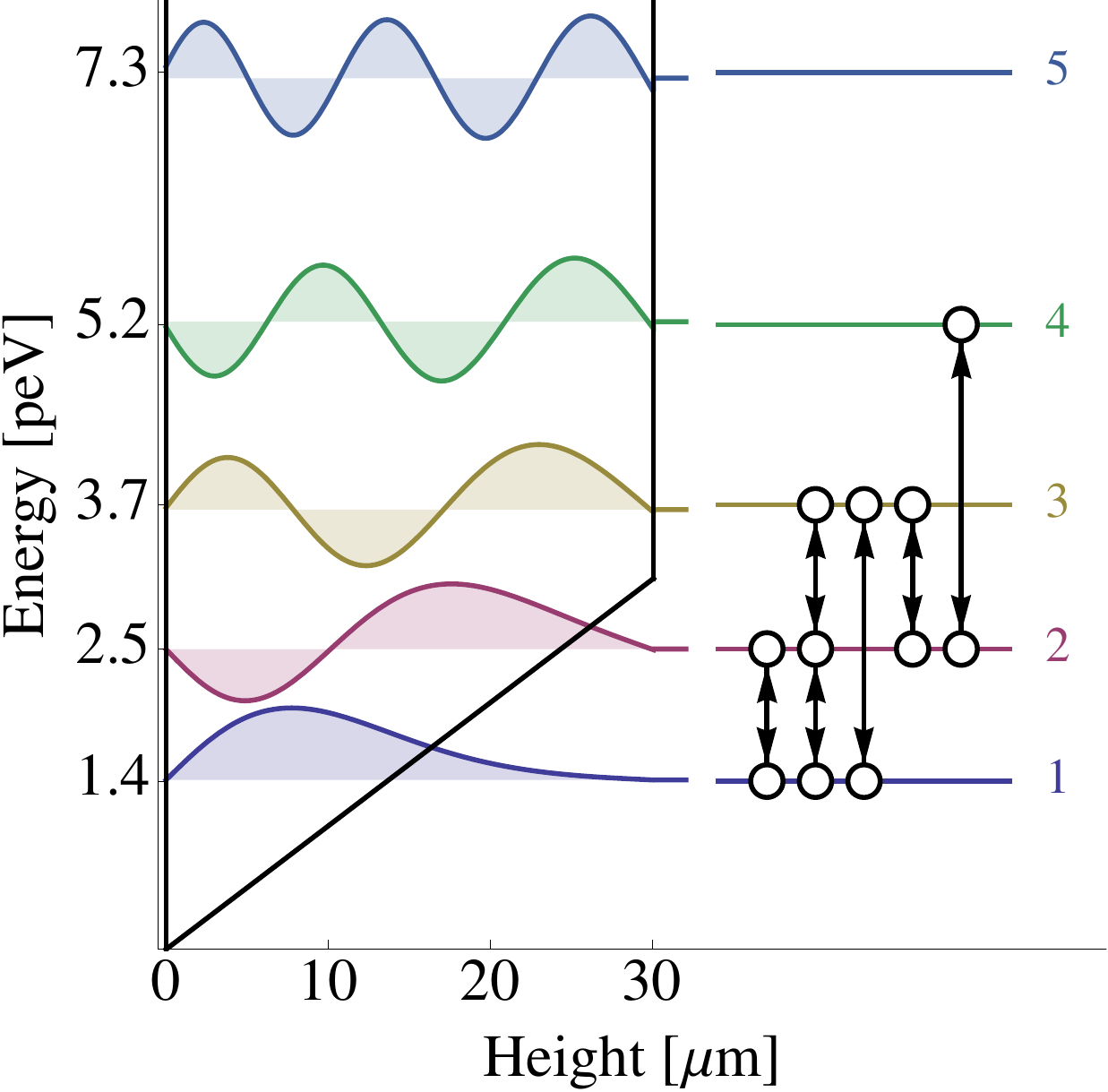}
	\includegraphics[width=4.6cm]{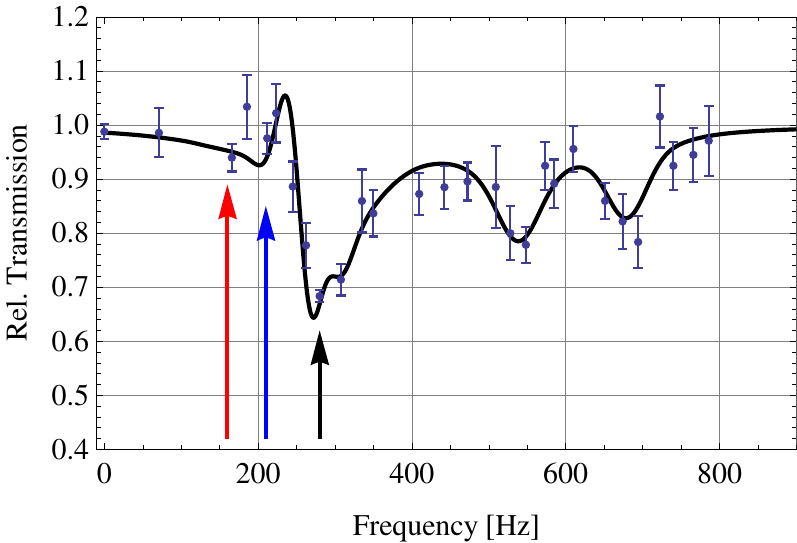}
\caption{Energy schema and results for the Gravity Resonance Spectroscopy:
Left: The lowest eigenstates and -energies with confining mirrors at bottom and top separated by 30.1~$\mu$m.
The observed transitions are marked by arrows.
Right: The transmission curve determined from the neutron count rate behind the mirrors as a function of oscillation frequency shows dips corresponding to the transitions shown on the left (see~\cite{Jenke2014}).\label{fig:GRS}}
\end{figure}

 Here we focus on the control and understanding of a gravitationally interacting elementary quantum system using the techniques of resonance spectroscopy. It offers a new way of looking at gravitation based on quantum interference. The system gives access to all parameters describing Newtonian gravitation: the mass and the distance. With the \emph{q}\textsc{Bounce} experiment, we examine gravity at short distances of microns. The energy scale is in the pico-eV range. In addition with the spin of the neutron, a hypothetical torsion coupling can be tested. Some authors believe that space-time continuum, viewed microscopically, should carry a torsion~\cite{Hehl85,Ivanov14}. The axion, a prominent dark matter candidate, would also provide a spin-mass coupling, which is examined. What is more, the Einstein equation has on the right hand side the energy-momentum tensor. In a modern interpretation, a scalar field, which might be responsible for the accelerated expansion of the universe, is mimicking such a vacuum energy, also called dark energy. With a hypothetical coupling to a scalar field, there are good prospects that neutrons solve the mysteries of dark energy. The inertial mass and the gravitational mass of the neutron will be determined from free fall alone and test the equivalence principle with quantum aspects. As a side effect, one can improve the limit on the charge of the neutron~\cite{durstberger2011probing}. The linear gravity potential leads to discrete, non-equidistant energy eigenstates as shown in Fig.~\ref{fig:GRS}, left, first measured in~\cite{Nesvizhevsky2002,Nesvizhevsky2005,Westphal2007} in a incoherent superposition. Evidence for states in a coherent superposition can be found in~\cite{Abele2009593c,Jenke2009318,jenke2013ultracold}.
The eigenenergies~$E_k$ are based on the slit width~$l$, the neutron mass~$m_n$, the reduced Planck constant~$\hbar$, and the acceleration of the earth~$g$.
As each transition can be addressed by its unique energy splitting, a combination of two states can be treated as a two-level system and resonance spectroscopy techniques can be applied~\cite{Jenke2011}. The quality of the neutron mirrors regarding roughness and waviness lead to systematic effects below 10$^{-19}$eV.

The ultra-cold neutron as a tool to resonance spectroscopy guaranties highest precision. It is insensitive to systematic effects due to electromagnetic ones plagued by other quantum objects like atoms or ions. An illustrative example is the calculation of the Casimir potential due to an atom at distance $r$ from an infinitely conducting surface. It can be written~\cite{Dim03} as
\begin{equation}
U_C=\frac{3\hbar c}{8\pi}\frac{\alpha_0}{r^4}.
\end{equation}
The conductivity is finite and corrections apply due to
and the dynamical polarizability of the atom. At length scales smaller than a wavelength
$\lambda/2\pi$, the
Casimir screening due to retardation becomes less effective
and the power law changes to $1/r^3$, the van
der Waals interaction. For $^{87}$Rb with $\alpha_0$ = 2.73$\times$10$^{-23}$ cm$^3$, at $r$ = 1$\mu$m, we get $U_C = 0.6 \mathrm{peV}$.

In contrast, a neutron has a small polarizability, so effects are expected many orders of magnitude below this scale.
The present experimental results allow us to search for any new kind of hypothetical gravity-like interaction at micron distances.
As shown in Fig.2, we address dark energy as a realization of quintessence theories in the so-called chameleon scenario~\cite{Khoury2004a,Mota2006,Mota2007,Waterhouse2006}, where a combination of the potential $V(\Phi,n)$ of a scalar field~$\Phi$ and a coupling~$\beta$ to matter together with model parameter~$n$ leads to the existence of an effective potential $V_{\rm eff}$ for the scalar field quanta, which depends on the local mass density~$\rho$ of the environment.

The experiment is most sensitive at $2 \leq n \leq 4$, where a chameleon interaction is excluded for $\beta > 5.8 \times 10^8$ (95\% C.L.).
The present limit from a fit to Fig.~\ref{fig:GRS}~\cite{Jenke2014} is five orders of magnitude lower than the upper bound from precision tests of atomic spectra~\cite{Brax2011}. The parameter space is restricted from both sides, as other experiments provide a lower bound of $\beta < 10$ at $n<2$~\cite{Adelberger2009,Brax2011}.

\bibliography{reference}

\end{document}